\documentclass[8.5pt,twoside,twocolumn]{article}
\usepackage[super,sort&compress,comma]{natbib} 
\usepackage{mhchem}
\usepackage{times,mathptm}
\usepackage{sectsty}
\usepackage{balance} 
\usepackage[usenames,dvipsnames]{color}

\usepackage{graphicx} 
\usepackage{lastpage}
\usepackage[format=plain,justification=raggedright,singlelinecheck=false,font=small,labelfont=bf,labelsep=space]{caption} 
\usepackage{fancyhdr}
\usepackage{amsmath,amssymb,amsfonts}

\begin{document}

\thispagestyle{plain}
\fancypagestyle{plain}{
\renewcommand{\headrulewidth}{1pt}}
\renewcommand{\thefootnote}{\fnsymbol{footnote}}
\renewcommand\footnoterule{\vspace*{1pt}%
\hrule width 3.4in height 0.4pt \vspace*{5pt}} 
\setcounter{secnumdepth}{5}

\makeatletter 
\def\subsubsection{\@startsection{subsubsection}{3}{10pt}{-1.25ex plus -1ex minus -.1ex}{0ex plus 0ex}{\normalsize\bf}} 
\def\paragraph{\@startsection{paragraph}{4}{10pt}{-1.25ex plus -1ex minus -.1ex}{0ex plus 0ex}{\normalsize\textit}} 
\renewcommand\@biblabel[1]{#1}            
\renewcommand\@makefntext[1]%
{\noindent\makebox[0pt][r]{\@thefnmark\,}#1}
\makeatother 
\renewcommand{\figurename}{\small{Fig.}~}
\sectionfont{\large}
\subsectionfont{\normalsize} 

\fancyfoot{}
\fancyfoot[C]{\footnotesize{\sffamily{\thepage~\textbar~\pageref{LastPage} }}}
\fancyhead{}
\renewcommand{\headrulewidth}{1pt} 
\renewcommand{\footrulewidth}{1pt}
\setlength{\arrayrulewidth}{1pt}
\setlength{\columnsep}{6.5mm}
\setlength\bibsep{1pt}

\newcommand{\ie}{\textit{i.e.}}
\newcommand{\eg}{\textit{e.g.}}
\newcommand{\F}{\mathcal{F}}
\newcommand{\x}{\mathbf{x}}
\newcommand{\HH}{\mathcal{H}}
\newcommand{\rmd}{{\rm d}}
\newcommand{\rme}{{\rm e}}
\newcommand{\reff}[1]{(\ref{#1})}
\newcommand{\BC}{BC}
\newcommand{\B}{{\mathcal B}}
\newcommand{\new}[1]{{\color{red} #1}}

\twocolumn[
 \begin{@twocolumnfalse}
\noindent\LARGE{\textbf{Critical Casimir forces steered by patterned substrates}}
\vspace{0.6cm}

\noindent\large{\textbf{Andrea Gambassi\textit{$^{a}$} and
S.~Dietrich\textit{$^{b,c}$}}}\vspace{0.5cm}

\vspace{0.6cm}

\noindent \normalsize{%
Among the various kinds of effective forces in soft matter, the spatial range and the direction of the so-called critical Casimir force -- which is generated by the enhanced thermal fluctuations close to a continuous phase transition -- 
can be controlled and reversibly modified to an uncommonly large extent. 
In particular, minute temperature changes of the fluid solvent, which provides the near-critical thermal fluctuations, lead to  a significant change of the range and strength of the effective interaction among the solute particles. This feature allows one to control, \eg, the aggregation of colloidal dispersions or the spatial distribution of colloids in the presence of chemically or topographically patterned substrates. The spatial direction of the effective force acting on a solute particle depends only on the surface properties of the immersed particles and can be spatially modulated by suitably patterned surfaces. 
These critical Casimir forces are largely independent of the specific materials properties of both the solvent and the confining surfaces. 
This characteristic universality of critical phenomena allows systematic and quantitative theoretical studies of the critical Casimir forces in terms of suitable representative and simplified models.
Here we highlight recent theoretical and experimental advances concerning critical Casimir forces with a particular emphasis on the numerous possibilities of controlling these forces by substrate patterns. 
}
\vspace{0.5cm}
\end{@twocolumnfalse}
 ]

\footnotetext{\textit{$^{a}$~SISSA -- International School for Advanced Studies and INFN, via Bonomea 265, 34136 Trieste, Italy. Tel: +39 040 3787 285; E-mail: gambassi@sissa.it}}
\footnotetext{\textit{$^{b}$~Max-Planck-Institut f\"ur Metallforschung, Heisenbergstr.~3, 70569 Stuttgart, Germany.}}
\footnotetext{\textit{$^{c}$~Institut f\"ur Theoretische und Angewandte Physik, Pfaffenwaldring 57, 70569 Stuttgart, Germany.}}

\section{Introduction}

The physical properties of soft matter can be described by a variety of  effective interactions of different nature involving, \eg, screened electrostatics, hydrodynamics, and depletion. These depend, inter alia, on the thermodynamic state of the system itself and on its structure at various length scales.    
Thermal fluctuations can also generate effective forces: this phenomenon~\cite{FdG} is analogous to the celebrated Casimir effect~\cite{Casimir} of quantum electrodynamics, in which two parallel, uncharged, metallic plates in vacuum perturb the spectrum of the quantum fluctuations of the electromagnetic field in the confined space so that the energy of the system depends on the separation $L$ between the plates. Accordingly, changing $L$ results in a variation of the energy and therefore an effective normal force acts on the plates. 

In the case we are interested in here, the relevant fluctuations are those of the spatially varying order parameter $\phi(\x)$ of a second-order phase transition occurring in a fluid medium and they are of thermal nature~\cite{gamb-rev}. Upon approaching the critical point of such a transition, the free energy  $\F$ of the medium is dominated by the thermal fluctuations of the order parameter $\phi$, which are spatially correlated across a typical distance $\xi$. This so-called correlation length represents the relevant length scale close to a critical point and it diverges algebraically at criticality. The quantity playing the role of the order parameter depends strongly on the specific phase transition of interest and indeed $\phi$ may be a quantity as diverse as the quantum wave-function of the superfluid in the case of the normal-to-superfluid transition of ${}^4$He or the difference $c_A(\x)-c_B(\x)$ between the local concentrations $c_{A,B}$ of two species $A$ and $B$ in the case of  a classical binary liquid mixture undergoing demixing. In spite of these different microscopic origins of $\phi$, the relevant singularities in the thermodynamic and structural properties of the medium appearing upon approaching the critical point can be inferred from the effective Hamiltonian $\HH[\phi]$, which renders the probability $\propto \rme^{-\HH[\phi]}$ of the fluctuations of the order parameter and therefore determines the effective free energy $\F = - k_BT\ln \int [\rmd\phi] {\rm e}^{-\HH[\phi]}$. Although a priori $\HH[\phi]$ depends on the specific material and microscopic properties of the medium and on its structure at various length scales, upon approaching the critical point only a few of these features turn out to determine the thermodynamic and structural properties  at mesoscopic length scales. This allows a drastic simplification of $\HH[\phi]$, the functional form of which is then dictated by  few gross features of the medium, such as the range and the symmetry of the microscopic interactions and general properties of the order parameter.  As a result, systems which differ significantly at the microscopic level can display the same singular behavior upon approaching a critical point -- a fact which gives rise to the notions of \emph{universality} and \emph{(bulk) universality classes} in critical phenomena.  
As an illustration, the correlation length $\xi$ as a function of the temperature $T$ increases algebraically as $\xi \simeq \xi_0|T/T_c-1|^{-\nu}$, where $\nu$ is a critical exponent, $T_c$ is the critical temperature, and $\xi_0$ is a molecular scale of the medium. Different from $T_c$ and $\xi_0$, the exponent $\nu$ is universal, as it does not depend on the material properties of the medium. 

Immersing a rigid object such as a plate or a particle into a critical fluid changes the fluctuations of the order parameter as a result of the interaction between the molecules of the object and those of the fluid. In particular, the perturbation produced by the object extends into the surrounding fluid within a typical distance $\xi$. Upon approaching the critical point, it turns out that the microscopic properties of the interaction between the object and the fluid are largely irrelevant for determining the thermodynamic and structural properties of this inhomogeneous system at a mesoscopic scale, leading to the notion of \emph{surface universality classes}~\cite{binder,diehl}. In fact, the presence of such objects can effectively be accounted for by suitable boundary conditions (\BC) for the order parameter $\phi$ at the surface of the object. 
In particular, the presence of the boundary can lead to a local enhancement of the order parameter $\phi$, such that either positive or negative values of $\phi$ are favored close to it: These adsorption preferences effectively lead to the \BC\ which we denote by $(+)$ or $(-)$, respectively. Alternatively, there might be no preferential adsorption, which results in the Dirichlet $(O)$ \BC\ $\phi=0$ at the surface. 
If a second object is immersed into the fluid it will modify the order parameter $\phi$ which, in turn, is de facto affected by the presence of the first object. This leads to an effective interaction between the two objects only if the minimal distance $L$ between them is smaller than the correlation length $\xi$. Indeed, if $L>\xi$, the second object is exposed to a fluid which is basically unperturbed by the first one. For $L<\xi$ an effective interaction between the two objects emerges with its spatial range set by the correlation length $\xi$. Accordingly, one can expect that immersing a third object sufficiently close to the previous two will also affect the mutual interaction of those previous ones, leading to non-additive many-body effects. Upon approaching the critical point, this effective interaction -- known as the \emph{critical Casimir effect} -- inherits the property of universality which characterizes the critical behaviors in the bulk and at surfaces, so that it depends only on the bulk and surface universality classes, the latter being characterized by the effective \BC\ at the surfaces, and possibly on the shape of the boundaries.  
For a fluid in the film geometry at temperature $T$, confined between two parallel plates of large transverse area $S$ and at a distance $L$ with total volume $V = S L$, the free energy $\F_V$ depends on $L$ and therefore each of the two confining plates is subject to a force $F_V = - \partial \F_V/\partial L = S p_b(T) + F_C + \mbox{corr.}$, where $p_b(T) \equiv F_V(L\rightarrow \infty)/S$ is the pressure which the \emph{bulk} fluid would exert on both sides of each single wall separately (\ie, for $L\rightarrow\infty$). The so-called critical Casimir force $F_C$ is  the leading $L$-dependent contribution to $F_V$ for  $L\gg\xi_0$ and it represents the \emph{effective interaction} between the two walls due to their simultaneous presence within the fluid.  Upon approaching the critical point, \ie, for $\xi\gg\xi_0$, the temperature dependence of  $F_C$ -- up to  non-singular background correction terms -- is encoded in a \emph{universal scaling function} $\vartheta$, along with its dependence on the effective \BC:
\begin{equation}
F_C  =  S\frac{k_BT}{L^3}\vartheta(L/\xi).
\label{eq:FC}
\end{equation}
Focussing now on the \emph{total} force on one of the two plates which is completely surrounded  by the fluctuating fluid, the term $S p_b(T)$ in $F_V$ is counterbalanced by the bulk force $- S p_b(T)$ exerted on the plate from the unconfined fluid outside of the film and therefore at leading order the total force is given by $F_C$ alone.
As anticipated, the range of $F_C$ is set by the correlation length $\xi$ of the critical fluctuations. The scaling function $\vartheta$ is universal in that it depends only on certain gross features of the critical behavior of the system and on the kind of \BC\ imposed by the two plates on the order parameter $\phi$, i.e., $(+,+)$ or $(+,-)$ for a binary liquid mixture of classical fluids and $(O,O)$ for pure $^4$He (because the superfluid order parameter does not couple to the substrate potential).  This universality allows one to determine the function $\vartheta$ for the force arising at the critical demixing point of classical binary mixtures by studying slabs of the Ising model with surface fields, which mimic $(+,+)$ and $(+,-)$ \BC.  On the same footing, for the superfluid transition in $^4$He  $\vartheta$  can be determined by studying  a slab of the so-called XY model with free \BC, realizing $(O,O)$ \BC. Recently these studies  have been carried out via Monte Carlo simulations~\cite{EPL-MC,MC-long,hucht} (see also Refs.\cite{hasen1,hasen2,hasen3,hasen4}). It turns out that the resulting force is attractive for equal \BC\ [$(+,+)$, $(O,O)$] and repulsive for unequal \BC\ [$(+,-)$, $(+,O)$\cite{mac-He4,He-mix-th}]. 

\section{Experimental evidence: from wetting films to colloids}
\begin{figure*}
\centering
  \includegraphics[height=4.3cm]{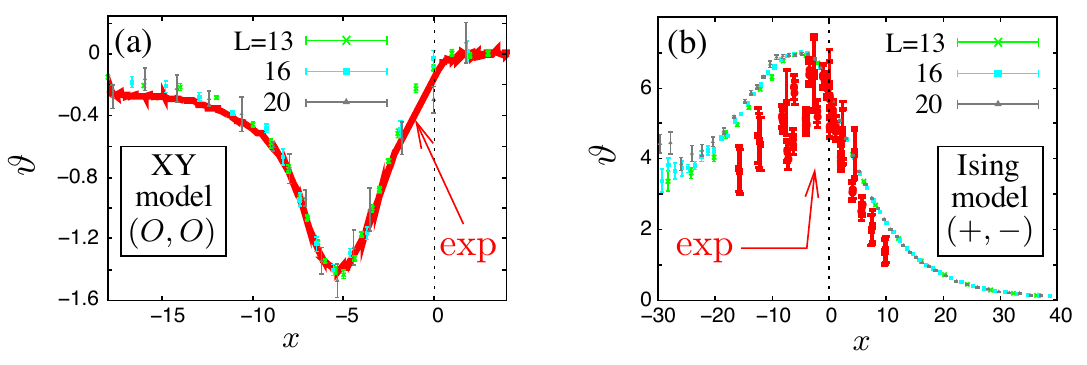}  
  \caption{Temperature dependence of the universal scaling function [see Eq.~\reff{eq:FC}] of the critical Casimir force in films of (a) $^4$He and (b) of a classical binary mixture, as a function of the scaling variable $x= (T/T_c-1)(L/\xi_0)^{-\nu}$.  Data points refer to Monte Carlo simulations of (a) the XY and (b) the Ising model in slabs of thickness $L$ with $(O,O)$ and  $(+,-)$ \BC, respectively\cite{EPL-MC,MC-long}. The agreement with the corresponding experimental data\cite{He4-exp-1,He4-exp-2,mixt-exp} ("exp") is remarkable. In panel (a), the slight but visible discrepancy between the asymptotic values approached by the experimental and by the Monte Carlo data for $x\lesssim -15$  is due to and almost completely accounted for by helium-specific surface fluctuations\cite{He4-lowT}.}
  \label{fig:indirect}
\end{figure*}

\begin{figure}[h]
\centering
  \includegraphics[height=5.5cm]{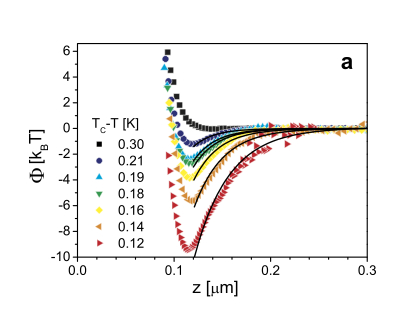}   
  \includegraphics[height=5.5cm]{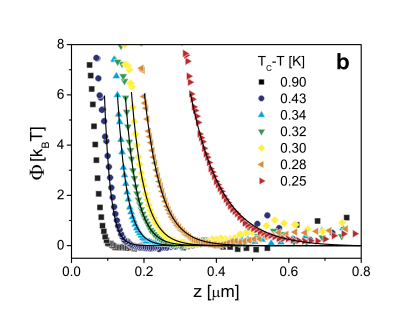}  
  \caption{Potential $\Phi(z)$ of the force acting on a colloid immersed in a water-lutidine mixture and at distance $z$ from an hydrophilic substrate,  for various temperatures  of the mixture\cite{Nature-Cas,Nature-long}.
Upon approaching the critical point, an attractive Casimir force is observed with an hydrophilic particle [$(-,-)$ \BC, panel (a)], whereas with an hydrophobic particle [$(+,-)$ \BC, panel (b)] the force is repulsive. The solid lines correspond to the theoretical predictions for the contribution $\Phi_C(z)$ of the critical Casimir force to the total potential $\Phi(z)$, without adjustable parameters\protect{\cite{Nature-long}}.}
  \label{fig:direct}
 \end{figure}

A rough estimate of $F_C$ reveals that for $T = T_c$ at room temperature, for two plates with $S=1\,\mbox{mm}^2$ and at a distance of $1\,\mu$m, the expected force is about $10^{-10}$N. Due to this weakness its experimental detection is particularly difficult. In addition, the experimental realization of a film of fluid with sufficiently accurate and  constant thickness $L$  poses a challenge due to the technical problems in realizing and maintaining the alignment between confining surfaces less than a micrometer apart. A solution to this problem -- which motivated subsequent experimental investigations -- is provided naturally by wetting films\cite{krech:92wet,indek}. If a bulk vapor phase, which is thermodynamically close to the condensation line, is exposed to a suitable substrate, a fluid wetting film of a certain thickness $L$ forms at the surface of the substrate. Strong substrates undergo complete wetting, so that $L$ becomes macroscopically large and diverges if the undersaturation of the vapor phase vanishes. In the resulting wetting film the fluid is naturally confined by the perfectly aligned fluid/substrate and the fluid/vapor interfaces. Far from $T_c$ its equilibrium thickness $L$ is determined by van-der-Waals dispersion forces and by the undersaturation. If such a confined fluid is driven thermodynamically towards a second-order phase transition, its critical fluctuations are confined within the film and therefore they give rise to a critical Casimir force $F_C$ as described by Eq.~\reff{eq:FC}. This force $F_C$ acts on the liquid/vapor interface and displaces it from its equilibrium position it would have under the influence of dispersion forces alone, i.e., in the absence of critical fluctuations. As a result $L$ itself changes due to the critical Casimir force. A detailed study of the temperature dependence of such a change allows the experimental determination of $F_C$ and of the associated scaling function $\vartheta$ in Eq.~\reff{eq:FC}. This kind of approach has provided the first indirect experimental evidence of critical Casimir forces in wetting films of pure liquid $^4$He~\cite{He4-exp-1,He4-exp-2} and $^3$He-$^4$He mixtures\cite{He-mix-exp} close to the superfluid transition as well as in wetting films of classical binary liquid mixtures\cite{mixt-exp,mixt-exp-Bonn}. In Fig.~\ref{fig:indirect} we compare the scaling functions determined in some of these experiments\cite{He4-exp-1,He4-exp-2,mixt-exp} with the corresponding theoretical predictions obtained by Monte Carlo simulations\cite{EPL-MC,MC-long} of suitable lattice models with appropriate \BC, as discussed above. (Effective mean-field theories provide a qualitative understanding of the shapes of these functions~\cite{mac-He4,He-mix-th,Krech:MFT,Zandi:MFT}, which turns quantitative if improvements are introduced\cite{BUp-1,BUp-2}.) The numerical data corresponding to different slab thicknesses $L$ fall onto a single master curve $\vartheta$ if plotted as a function of the proper scaling variable $x= (T/T_c-1)(L/\xi_0)^{-\nu}$, where the values of the critical exponents are those characteristic for the phase transition investigated, i.e., $\nu\simeq 0.66$ in panel (a) and $\nu\simeq 0.63$ in panel (b). For the normalization of the abscissa, the non-universal amplitude $\xi_0$ takes different values for the numerical and the experimental data [both in (a) and (b)], as independently inferred from the temperature dependence of  the corresponding bulk correlation lengths.  With these normalizations, there are no adjustable parameters left, providing a remarkable good agreement between the theoretical and the experimental data. Note that the resulting critical Casimir force is attractive in (a) but repulsive in (b).  
The direct measurement of the Casimir force becomes possible if one considers the force acting on a spherical particle, immersed in the fluid, when it approaches the wall of the container of the critical fluid. As in the case of parallel plates, both the wall and the sphere impose effective \BC\ on the order parameter $\phi$ and therefore a critical Casimir force $F_C$ acts on the sphere. If the distance of closest approach $z$ between the sphere and the wall is much smaller than the radius $R$ of the particle, the potential $\Phi_C(z)$ of the critical Casimir force takes the form
\begin{equation}
\frac{\Phi_C(z)}{k_BT} = \frac{R}{z}\vartheta_{\circ|}(z/\xi)
\label{eq:PhiC}
\end{equation}
where the function $\vartheta_{\circ|}$ can be expressed in terms of $\vartheta$\cite{Nature-Cas,Nature-long} and shares with it the qualitative and universal properties. 
Using Total Internal Reflection Microscopy\cite{Prieve} with its femto-Newton force resolution 
it has been possible\cite{Nature-Cas,Nature-long} to study the onset of critical Casimir forces acting on a single micrometer-sized colloid near a wall as the temperature $T$ of a water-lutidine solvent is increased towards its (lower) critical value $T_c\simeq 34^\circ\mbox{C}$ at fixed critical lutidine mass fraction.
We first consider the case of an hydrophilic colloid and a hydrophilic wall, corresponding to $(-,-)$ \BC. (With reference to the water-lutidine mixture, "$+$" and "$-$" indicate the preferential adsorption of lutidine and water, respectively.)  Far from the critical point the measured potential $\Phi(z)$ (apart from the contribution of buoyancy, which is easily subtracted from the data presented in Fig.~\ref{fig:direct}) is given solely by the electrostatic repulsion between the colloid and the wall, as shown in Fig.~\ref{fig:direct}(a) for $T_c-T=0.3\,\mbox{K}$. (In the present experimental conditions, van-der-Waals forces turn out to be negligible\cite{Nature-long}.) However, upon heating the mixture, an increasingly deep potential well develops, which is due to the attractive Casimir force providing a negative 
contribution $\Phi_C(z)$. As expected theoretically, the spatial range $\xi$ of $\Phi_C(z)$ increases upon increasing $T$ towards $T_c$. The set of measured potentials $\Phi_C(z)$ can be compared with the corresponding theoretical prediction [Eq.~\reff{eq:PhiC}, solid lines in Fig.~\ref{fig:direct}(a)], resulting in a remarkable agreement.  
If one exchanges the hydrophilic colloid for a hydrophobic one, the corresponding \BC\ change from $(-,-)$ to $(-,+)$ and on theoretical grounds one expects repulsion. As before, far below the critical point, the interaction potential consists of the electrostatic repulsion only, as in Fig.~\ref{fig:direct}(b) for $T_c-T=0.9\,\mbox{K}$. Upon heating the mixture, the repulsive part of the potential curves shifts towards larger values of $z$ due to the fact that, as expected, repulsive Casimir forces are acting on the particle and yield a positive contribution $\Phi_C(z)$ to the total potential $\Phi(z)$. As in the case of Fig.~\ref{fig:direct}(a), the agreement with the corresponding theoretical prediction  [Eq.~\reff{eq:PhiC}, solid lines in Fig.~\ref{fig:direct}(b)] is remarkable. By changing in addition the adsorption preference of the wall from hydrophilic $(-)$ to hydrophobic $(+)$ via a suitable chemical surface treatment, attraction is recovered\cite{Nature-long}.  

These experiments demonstrate that the critical Casimir force can be put to work in soft matter systems, in which the uncommonly high degree of tunability of the force both with respect to strength and character (attractive/repulsive) might be harnessed to serve dedicated purposes for, \eg, colloidal suspensions. 

\section{Patterned substrates}
The cases discussed above concerning the film geometry and a spherical colloid in front of a flat substrate are characterized by a critical Casimir force which is directed perpendicularly to the flat surface(s), due to the high spatial symmetry of the problem resulting from the homogeneity of the confining surfaces. However, a \emph{lateral} component of the critical Casimir force emerges as soon as this symmetry is reduced. For instance, this is the case for the film geometry if both surfaces are suitably patterned -- either chemically\cite{Casimir-patt-plates} or geometrically\cite{Casimir-geo} -- or for a homogeneous colloid in front of a patterned surface\cite{EPL-pattern,long-pattern}. A chemically patterned surface is characterized by a spatially modulated adsorption preference which, in turn, induces spatially varying effective boundary conditions for the order parameter $\phi$.
The resulting lateral component of the force allows one to additionally control the spatial direction of the total critical Casimir force, which can also be locally modified along the substrate.

In order to study theoretically the effects of inhomogeneous surfaces on the critical Casimir force, three different and partly complementary approaches can be followed: 
(a) Numerical minimization of the Landau-Ginzburg effective energy  $\HH$, which yields the spatially varying mean-field order parameter profile $\bar\phi(\x)$ from $\delta\HH[\phi]/\delta\phi|_{\phi=\bar\phi}=0$ where $\bar\phi$ satisfies the proper effective \BC\ at the surfaces (either flat or curved). In terms of $\bar\phi$ the effective free energy  functional $\F$ is given by $\F = k_B T \HH[\bar\phi]$, from which the critical Casimir force follows.
While this approach properly accounts for the geometry of the boundaries and, in four spatial dimensions, also for the critical behavior of the system,  in three dimensions it provides only an approximation of this behavior. 
(b) Derjaguin approximation (DA), which expresses the critical Casimir force acting on inhomogeneous or/and curved surfaces in terms of the force acting on homogeneous flat surfaces\cite{EPL-pattern,long-pattern}.
Within this approach, the geometry of the boundaries is accounted for only approximately, and indeed the accuracy of the DA improves if the confining surfaces are almost flat and homogeneous on the scale of the distance from each other -- which might be the case in experimentally relevant conditions.  
On the other hand, it is possible to take advantage of the quantitatively reliable  knowledge of the film scaling function $\vartheta$ in spatial dimension $d=3$ as obtained, \eg, from Monte Carlo simulations.  The range of geometrical parameters within which the DA provides accurate results can be assessed in $d=4$: the approach (a) allows one  to determine numerically the critical Casimir force (without any further approximation), which can also be calculated via DA as discussed above on the basis of the analytical knowledge of $\vartheta$ in $d=4$. We expect that this range of applicability of the DA carries over to $d=3$.  
(c) Monte Carlo simulations of lattice models with suitable geometry and \BC\ which, apart from the common numerical issues  inherent in numerical simulations (such as statistical uncertainties and finite-size effects), provide quantitatively accurate predictions for the critical Casimir potential in $d=3$. However, due to the difficulty to implement curved boundaries on a lattice structure, this technique is primarily applied to study films with chemically patterned boundaries.

The simplest chemical substrate pattern is obtained by joining two half-planes with different adsorption preferences, so that the effective \BC\ imposed on the order parameter $\phi$ changes steplike along a straight line (say, at $x=0$) while being homogeneous in the remaining lateral direction. If such a substrate is used together with a homogeneous one for confining a near-critical fluid in the film geometry, the geometrical symmetry of the problem  still prevents the emergence of an overall lateral critical Casimir force $F_C^\|$, while it allows the emergence of a critical Casimir torque parallel to the chemical step.
For such a system the influence of the presence of the chemical step on one surface is encoded in the dependence of the normal critical Casimir force $F_C^\bot$ on the aspect ratio $L/\sqrt{S}$ of the slab~\cite{PTD}.
Analogously, for any patterned substrate $F_C^\|=0$ if the second confining substrate is homogeneous.
(However, this is no longer the case if one is chemically/geometrically patterned and the other is curved, even if homogeneous, see below.) 
Instead,  $F_C^\|\neq 0$ if the adsorption preferences of both substrates change periodically along, say, the  $x$-direction, \eg, alternating between $(+)$ and $(-)$ on adjacent stripes of widths $L_+$ and $L_-$, respectively.  
For fixed distance $L$, the patterned plates are expected to align laterally such that each stripe on one of the plates faces a stripe with the same adsorption preference on the other plate, due to the attractive interaction between homogeneous plates having the same \BC.  This configuration corresponds to the minimum of the critical Casimir potential $\Phi_C$. If one of the two plates is displaced laterally by an amount $\delta$ from such a perfectly aligned position, a lateral critical Casimir force $F^{\|}_C$ emerges which tends to move the plates back to this aligned configuration. In addition to its dependence on $L$ and $\xi$, $F^{\|}_C$ as well as $F^{\bot}_C$ depend additionally on the displacement $\delta$ (with a period $P=L_++L_-$) and on the stripe widths $L_+$ and $L_-$. Upon approaching the critical point, such a dependence takes a scaling form\cite{Casimir-patt-plates} analogous to Eq.~\reff{eq:FC}, in which the rhs depends additionally on the dimensionless scaling variables $\delta/L$, $L_-/L_+$, and $L_+/\xi$. These scaling functions have been studied in detail\cite{Casimir-patt-plates} within the approach (a) mentioned above for the case of patterns with $(\pm)$ \BC. 
Note that in all the cases mentioned here, the spatial direction in which the resulting critical Casimir force $\vec{F_C} = \vec{F}_C^{\bot} + \vec{F}_C^{\|}$ acts depends on the geometrical features of the patterns and, due to the non-trivial dependence on the correlation length $\xi$, can be controlled to a large extent by changing the temperature.

Micro- and nanoscale devices, such as micro-electromechanical systems (MEMS), involve chemically homogeneous substrates which are topographically structured for dedicated purposes. These substrates, when used in order to confine a near-critical fluid, are generally subject to lateral critical Casimir forces. %
Consider, \eg, the case of two identical substrates with a sawtooth periodic structure of grooves and ridges along the $x$ direction (while being translationally invariant in the remaining direction) which face each other at a fixed minimal surface-to-surface distance $L_{\rm tip}$. Since the critical Casimir force between two homogeneous planar substrates varies strongly with the distance between them, the lateral equilibrium position of the substrates corresponds to the ones for which the distance between the ridges on the two opposing substrates is minimal, such that they face each other. A lateral displacement from such a configuration results in a lateral critical Casimir force $F^{\|}_C$ which tends to bring the system back into one of the equilibrium configurations, as it has been investigated within the approach (a) discussed above for a family of regular, periodic patterns\cite{Casimir-geo}.  Also in this case, the dependence of both the normal and the lateral critical  Casimir forces $F_C^{\|,\bot}$ takes a scaling form in terms of suitable dimensionless variables quantifying the relative displacement and describing the topographic patterns on the substrates, such as the opening angle $\gamma$ of the grooves as well as the ratios $L_{\rm tip}/\xi$ and $h/\xi$ between $L_{\rm tip}$ and the groove depth $h$, respectively, and the correlation length $\xi$.  Within the similar geometrical set-up of sinusoidally corrugated plates, an analogous effect has been observed both theoretically and experimentally\cite{patt_CasQED_th,patt_CasQED_ex} involving the long-ranged force associated with the fluctuations of the electromagnetic field in vacuum.

In addition to their relevance for many technological applications, colloidal particles are widely used as model systems and for the investigation of soft matter properties. It is therefore important to be able to control and possibly modify the forces which act on them when dispersed in a solvent. Lateral critical Casimir forces  naturally provide such an opportunity: In the simplest instance of two flat substrates it is necessary that both involved surfaces are chemically patterned in order to obtain a lateral critical Casimir force, whereas this is no longer the case if one of the two surfaces is the curved one of a colloid. Accordingly, colloids with a homogeneous surface and exposed to a patterned substrate experience a critical Casimir force $\vec{F}_C$ with two nonzero vector components.. Its spatial direction can be finely controlled by changing temperature and it can be modulated in space by suitably patterning the surface either topographically or chemically. 
These forces can be used to generate confining potentials for colloids which can be switched on and off by ramping up or down the temperature of the near-critical solvent. This might find application for controlling the spatial distribution of colloids and possibly their rheology.  
The theoretical predictions for the potential\cite{EPL-pattern,long-pattern} of the forces acting on a single colloid in front of a patterned substrate can be obtained within approach (b) outlined above, taking advantage of the numerical knowledge of the scaling function $\vartheta$ in $d=3$. 

\begin{figure}[h]
\centering
  \includegraphics[width=0.8\columnwidth]{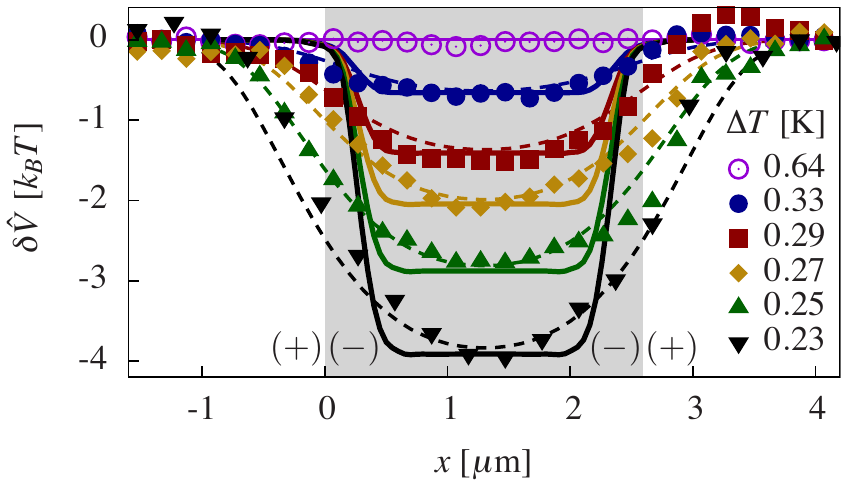} 
  \caption{Temperature-dependent effective potential $\delta\hat V(x)$ of a colloid with $(-)$\BC\ exposed to a substrate with $(+)$ \BC\ with a stripe with $(-)$ \BC. Symbols and lines correspond to experimental data\cite{Cas-patt-exp} and theoretical predictions\cite{EPL-pattern}, respectively (see the main text for details).}
  \label{fig:stripe}
\end{figure}

As an example, in Fig.~\ref{fig:stripe} we consider the behavior of a colloid with $(-)$ \BC\ and radius $R= 1.2\,\mu$m  immersed in a near-critical water-lutidine mixture, facing a substrate with $(+)$ \BC\ on which a stripe of width $L=2.6\,\mu$m and $(-)$ \BC\ has been realized by exposure to a focused ion beam\cite{Cas-patt-exp}. (In the following $x$ indicates the distance of the center of the colloid measured along the substrate from one edge of the stripe, whereas $D$ is the surface-to-surface distance.) If a dilute solution of colloids is exposed to such a substrate, due to the spatially varying total  potential $\Phi$ they distribute inhomogeneously in space with a number density $\rho \propto \rme^{-\Phi/(k_BT)}$. Via optical video microscopy one can determine the average projected areal density $\rho_{P} = \int_0^\infty \!\rmd D\, \rho$ and the associated effective potential $\delta \hat V$ defined (up to an irrelevant constant) such that $\rho_P \propto \rme^{-\delta \hat V/(k_BT)}$. The data points in Fig.~\ref{fig:stripe} correspond to the experimental ones for various values of the distance $\Delta T$ from the critical point, \ie, for various correlation lengths $\xi$. Far from the critical point, $\delta\hat V$ is de facto spatially constant and therefore the colloids are homogeneously distributed. Upon reducing $\Delta T$, the critical Casimir force sets in and the colloidal particles are attracted both vertically and laterally by the stripe, which shares the same \BC, while they are repelled from the remaining part of the substrate.  As a result a potential minimum develops and it can become so deep that effectively the colloid cannot escape from it, leading to a localization of the particles above the stripe. 
The theoretical predictions for the corresponding effective potential $\delta\hat V$ (which also takes into account the presence of the screened electrostatic interaction between the colloid and the substrate) are indicated by the solid and the dashed lines. The former assume that the stripe has straight boundaries in the direction $y$ perpendicular to $x$, whereas the latter accounts for the possibility that they  wiggle as a function of $y$ around their average position, a feature that could not be tested experimentally by independent methods. The remarkable agreement between the dashed lines and the experimental data underscores that critical Casimir forces are very sensitive to fine details of the patterned substrate and can in principle be used to resolve features which are not accessible otherwise. 

In the presence of homogeneous substrates, the attractive versus repulsive character of the critical Casimir force $F_C^\bot$ experienced by a colloidal particle is controlled by the \BC\  only and does not change with the surface-to-surface distance $D$. (Such a change can be observed in the \emph{total} force due to additional competing contributions of buoyancy, electrostatic, or van-der-Waals forces.)
\begin{figure}[h]
\centering
  \includegraphics[width=0.7\columnwidth]{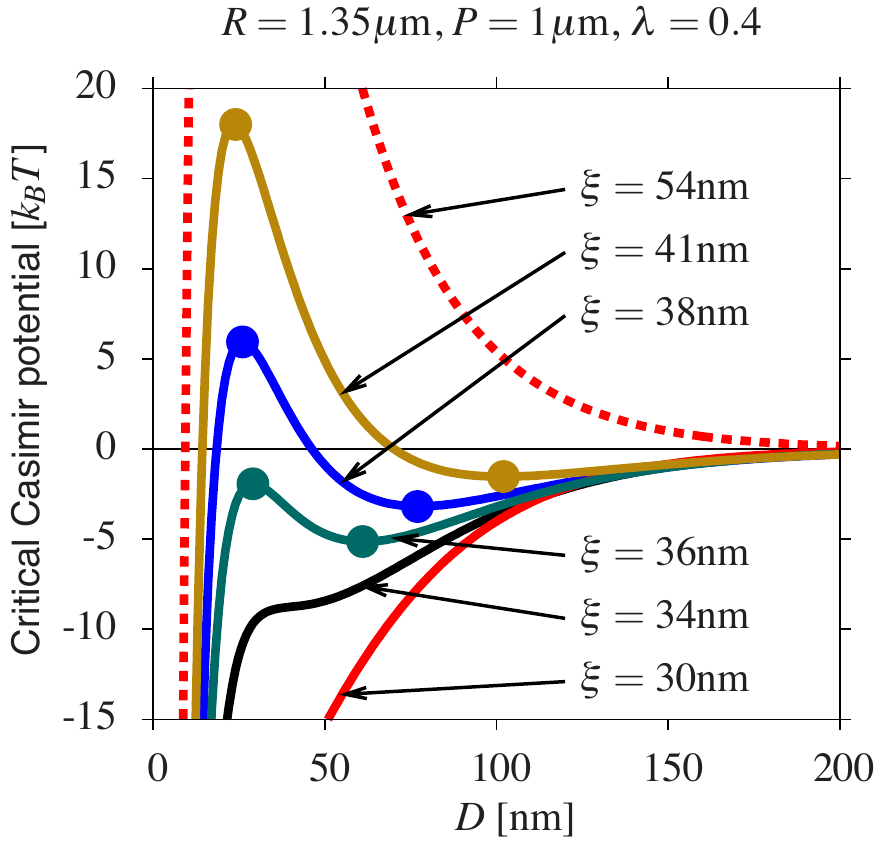}   
  \caption{Critical Casimir potential of a colloid of radius $R$ in lateral equilibrium close to a periodically patterned substrate as a function of $D$ and for various values of $\xi$.}
  \label{fig:levit}
\end{figure}
Instead, patterned substrates may induce this latter feature, which is clearly illustrated by the theoretical predictions\cite{long-pattern} reported in Fig.~\ref{fig:levit} for the critical Casimir potential. These predictions have been obtained within the approach (b), for the case of a colloid of radius $R=1.35\,\mu$m and, say, $(-)$ \BC, facing a substrate with a periodic pattern of period $P=1\,\mu$m constituted by alternating stripes with $(+)$ and $(-)$ \BC\ and different widths $L_{(\pm)}$, such that $\lambda \equiv L_{(-)}/P = 0.4$. Due to the structure of the pattern, the colloid is in lateral equilibrium only if the perpendicular projection of its center onto the substrate falls onto the midaxis of the stripe with $(-)$ \BC. The potential reported in Fig.~\ref{fig:levit} is calculated for this lateral position. If the particle is  pushed vertically almost to contact with the substrate ($D=0$), the normal force $F_C^\bot$ is dominated 
by the attractive contribution originating from the interaction between the colloid and that portion of the substrate with the same \BC. Upon increasing the distance $D$, the repulsive contribution due to those portions of the substrate with $(+)$ \BC\ becomes increasingly important. Depending on the relative strengths of these two opposing contributions -- controlled by $\lambda$ and by the correlation length $\xi$ --  at large distances $F_C^\bot$ can be repulsive ($\xi = 54\,$nm in Fig.~\ref{fig:levit}) or attractive ($\xi = 30\div 38\,$nm). In the latter case a  stable mechanical equilibrium can appear at a distance $D_s$, which is separated from the global minimum of the potential at $D=0$ by a significant potential barrier. 
This local minimum of the potential can be used to realize and control the levitation of a colloid close to a substrate, \emph{independently} of the possible action of additional forces, the ranges and strengths of which can be suitably controlled by a proper selection of the employed materials. The position $D_s$ of stable levitation is strongly affected by temperature via the correlation length $\xi$, with a response $\Delta D_s/\Delta T$ which can be as large as  $\simeq 500\,\mbox{nm}/\mbox{K}$ (for Fig~\ref{fig:levit}). $D_s(T)$ reaches macroscopic values for $T\rightarrow T_s\neq T_c$.

\section{Perspectives and applications}
There are numerous aspects of the critical Casimir force which call for further investigations, both for their conceptual and practical relevance. Among them are equilibrium properties of colloidal dispersions\cite{coll-Cas-1,coll-Cas-2,colloidal-agg-ex,aggr-comm}. In particular, we mention here the role of geometry\cite{Cas-nonspher}, of the effective boundary conditions\cite{Mohry,Bech-grad}, and the dynamical behavior of the force\cite{dyn-film,dyn-cas,Dean-cas-lett}. For example, non-spherical colloids are expected to experience a critical Casimir torque when approaching a flat substrate\cite{Cas-nonspher}. 
Such a torque could in principle find applications for controlling the orientation of macromolecules in near-critical solutions and their alignment with suitably patterned substrates. Analogously, critical fluctuations within the solvent can also be used to induce an orientation-dependent effective interaction between chemically patterned colloids (\eg, Janus particles) which could induce the formation of a variety of different structures. Weak adsorption preferences of the surfaces can cause rich cross-over phenomena between different effective \BC, which include the possibility of a change in the attractive versus repulsive nature of the critical Casimir force, even between two flat and homogeneous substrates, upon changing the temperature\cite{Mohry}. 
In addition to the time-independent and equilibrium properties mentioned above, critical Casimir forces can  display interesting dynamical and non-equilibrium aspects of universal nature\cite{dyn-film,dyn-cas,Dean-cas-lett}, which could find applications for controlling dynamical processes in colloidal suspensions.

\vspace*{4mm}
\noindent {\bf Acknowledgments}\\
AG is supported by MIUR within the program "Incentivazione della  
mobilit\`a di studiosi stranieri e italiani residenti all'estero".

\footnotesize{

\providecommand*{\mcitethebibliography}{\thebibliography}
\csname @ifundefined\endcsname{endmcitethebibliography}
{\let\endmcitethebibliography\endthebibliography}{}

}

\end{document}